\documentclass[final,5p,times,twocolumn]{elsarticle}

\usepackage{lineno}

\usepackage{graphicx}
\usepackage{amssymb}
\usepackage{subfigure}

\journal{NIM-A Proceedings (Elba 2012)}

\begin{document}

\begin{frontmatter}

\title{Novel Silicon n-in-p Pixel Sensors for the future ATLAS Upgrades}

\author[A]{A. La Rosa\corref{cor1}}
\ead{alessandro.larosa@cern.ch}
\author[B]{C. Gallrapp}
\author[C]{A. Macchiolo}
\author[C]{R. Nisius}
\author[B]{H. Pernegger}
\author[D]{R.H.  Richter}
\author[C]{P. Weigell}

\address[A]{Section de Physique (DPNC), Universit\'e de Gen\`eve, 24 quai Ernest Ansermet, Gen\`eve 4,  CH-1211, Switzerland}
\address[B]{CERN, Geneva 23, CH-1211, Switzerland}
\address[C]{Max-Planck-Institut f\"ur Physik (Werner-Heisenberg-Institut) F\"ohringer Ring 6, D-80805 M\"unchen, Germany}
\address[D]{Max-Planck-Institut Halbleiterlabor, Otto Hahn Ring 6, D-81739 M\"unchen, Germany}

\cortext[cor1]{Corresponding author.}


\begin{abstract}
\indent In view of the LHC upgrade phases towards HL-LHC the ATLAS experiment plans to upgrade the Inner Detector with an all silicon system.
The n-in-p silicon technology is a promising candidate for the pixel upgrade thanks to its radiation hardness and cost effectiveness, that allow for enlarging the area instrumented with pixel detectors.\\
\indent We present the characterization and performance of novel n-in-p planar pixel sensors produced by CiS (Germany) connected by bump bonding to the ATLAS readout chip FE-I3.
These results are obtained before and after irradiation up to a fluence of  $10^{16}$\,1-MeV\,n$_{\mathrm{eq}}$cm$^{-2}$, and prove the operability of this kind of sensors in the harsh radiation environment foreseen for the pixel system at HL-LHC. We also present an overview of the new pixel production, which is on-going at CiS for sensors compatible with the new ATLAS readout chip FE-I4.\\
\end{abstract}

\end{frontmatter}

\section{Introduction}
\label{sec:intro}
With the LHC collecting data at 8 TeV, plans are already advancing for a series of upgrades leading eventually to achieve five times the LHC design luminosity around 2022 in the so-called High Luminosity LHC (HL-LHC) project. Considering that the current Inner Detector (ID) would become inefficient during the expected HL-LHC operation scenario, a replacement of the full ID is foreseen in the long shutdown preceding the Phase II of HL-LHC. The future ATLAS tracker will consist of an all silicon-based system with new detector technologies. In this scenario the innermost layers of the ATLAS vertex detector system will have to sustain very high integrated fluences of more than $10^{16}$\ 1-MeV\,n$_{\mathrm{eq}}$cm$^{-2}$ with 3000\,fb$^{-1}$ total integrated luminosity at the end of the LHC lifetime around 2030.\\
\indent The n-in-p silicon technology is a promising candidate for the pixel upgrade thanks to its radiation hardness and cost effectiveness, that allows for enlarging the area instrumented with pixel detectors. The n-in-p pixel sensors discussed in this paper have been produced at CiS (Erfurt, Germany) with a geometry compatible with the ATLAS Pixel readout chip (FE-I3) \cite{FEI3}, in the framework of a common CERN RD50 Collaboration production. The interconnection between sensors and readout chips has been done at IZM (Berlin, Germany).

\section{Sensor description}
\label{sec:sensor}

The n-in-p pixel sensors have been produced by CiS on 4" wafer of high resistivity Float zone (Fz) material, with a thickness of 300\,$\mu$m. 
This sensor technology requires a single-sided process, which implies a reduced number of process steps and leads to a cost reduction. Thanks to the absence of bulk type inversion and to the fact that the main junction is always between the n$^{+}$ implanted pixels and the p-type bulk, the guard ring structure can be placed on the front-side while the back-side is implanted with a uniform p$^{+}$ implantation. To achieve a narrow inactive region two different guard-rings structures with different widths, have been implemented, with one of them characterized by a reduced non-active area with respect to the 1\,mm per side of the ATLAS Pixel sensors \cite{atlaspixel}.
To prevent sparks between the area outside the guard ring and the readout chip a  BCB (BenzoCycloButene, Cycloten) layer has been applied to the pixelated side of the n-in-p pixel sensors.
More details about sensor design and process are given in \cite{NIM_CiS_ALR}.

\section{Results}
\label{sec:results}

\begin{figure}[h!]
\centering
\includegraphics[scale=0.35]{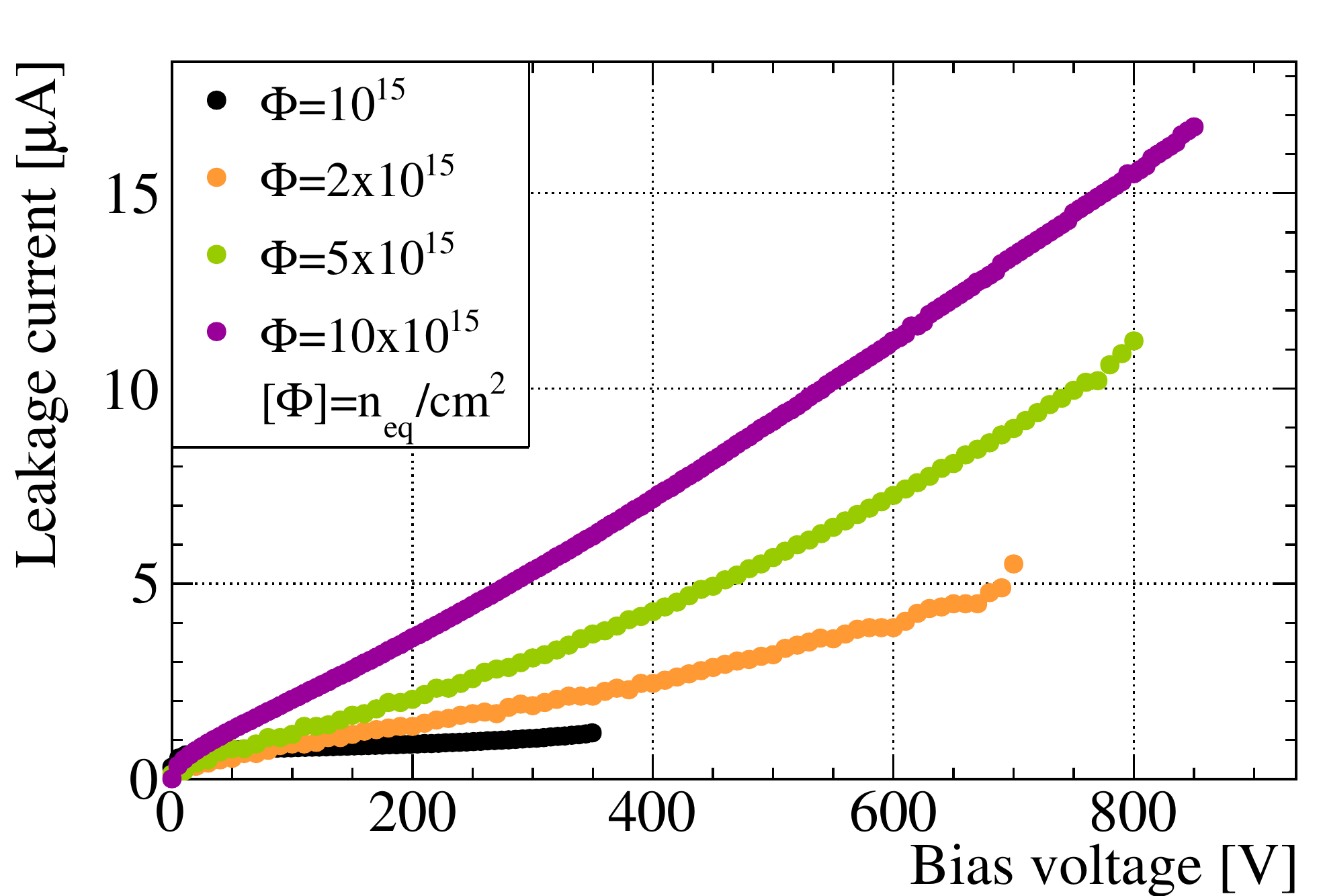}
\caption{IV characteristics of the selected modules irradiated up to 1, 2, 5 and 10\,x\,10$^{15}$\,1-MeV\,n$_{\mathrm{eq}}$cm$^{-2}$ respectively. All the measurements have been scaled to a temperature of $-20\,^{\circ}\mathrm{C}$. }
\label{fig:IVirrad}
\end{figure}

\begin{figure}[h!]
\centering
\includegraphics[scale=0.4]{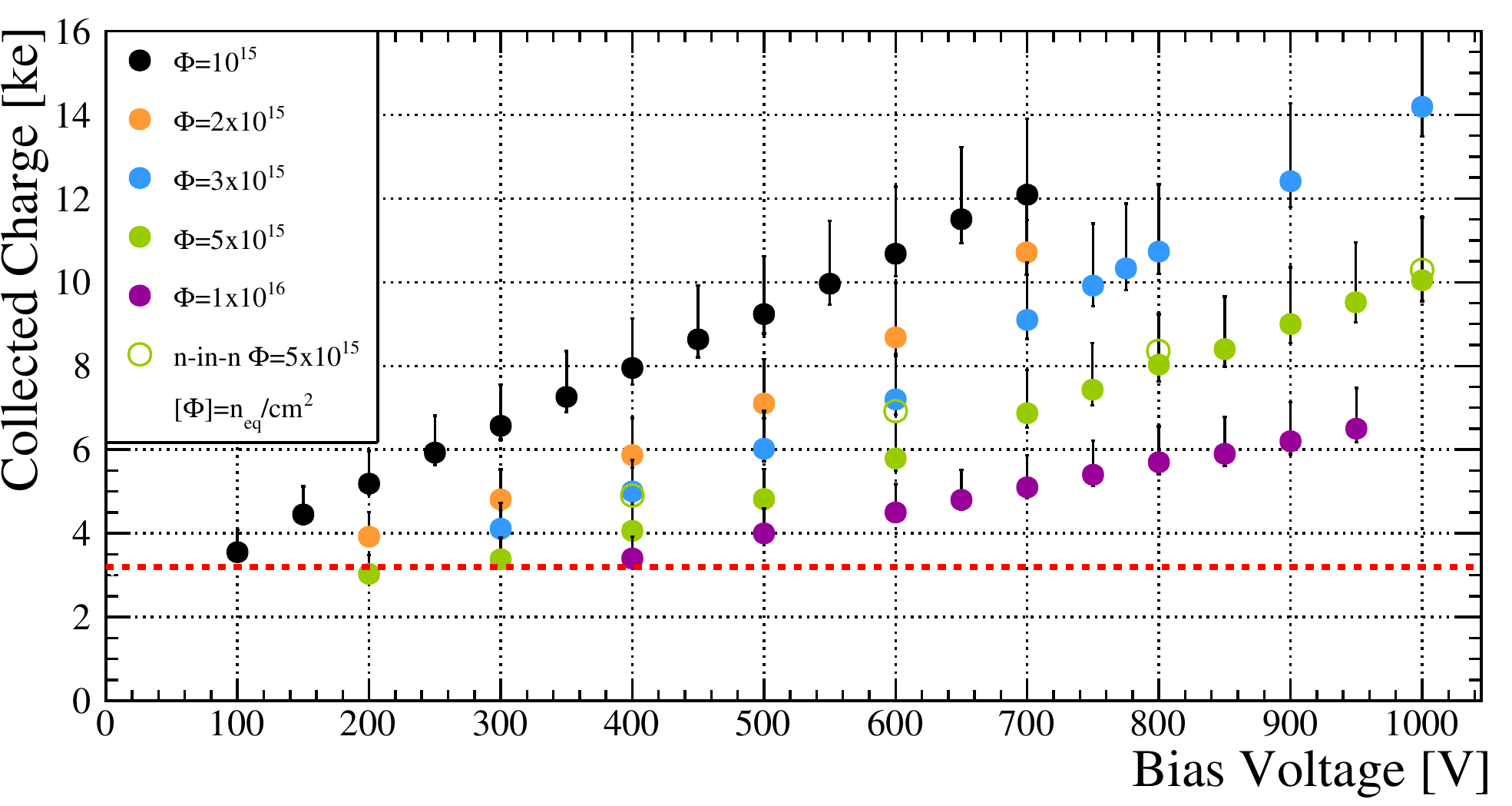}
\caption{Charge collection by modules irradiated up to $10^{16}$\,1-MeV\,n$_{\mathrm{eq}}$cm$^{-2}$ as a function of the bias voltage with front-end chips tuned to a threshold of 3.2\,ke. Results from the n-in-n modules are from Ref. \cite{NIM_Do}.}
\label{fig:MPVall}
\end{figure}

\indent To investigate the performance and the fluence range in which the n-in-p sensor technology can be used, the modules have been characterized by measuring the leakage current, the noise and the response to radioactive sources in the laboratory. Then, to measure the tracking efficiency, the charge sharing probability, and the charge collection, the modules were also tested at the CERN SPS with a 120 GeV/c $\pi$$^{+}$ beam.  
A selection of the most representative results, obtained before and after irradiation up to a fluence of $10^{16}$\,1-MeV\,n$_{\mathrm{eq}}$cm$^{-2}$, are reported in this section.\\
\indent {\bf Leakage current.} Before irradiation, all the manufactured modules show leakage currents below 0.6\,$\mu$A, when operated at a bias voltage of 150\,V \cite{NIM_CiS_ALR}. As expected, after irradiation up to $10^{16}$\,1-MeV\,n$_{\mathrm{eq}}$cm$^{-2}$, the breakdown voltages shifted to higher values and for the highest fluence it exceeds 800\,V. 
When scaled to the same temperature, the leakage currents increase with the fluence and the IV characteristics of selected modules are shown in Figure \ref{fig:IVirrad}.
 \\
\indent {\bf Charge collection.} The charge collection has been measured using a $^{90}$Sr $\beta$-source keeping the modules
 in a climate chamber at low humidity and stable air temperature ($+20\,^{\circ}\mathrm{C}$ for not irradiated and between $-60\,^{\circ}\mathrm{C}$ and $-20\,^{\circ}\mathrm{C}$ for irradiated modules). Prior to the source tests, the front-end chip has been tuned to a threshold of 3.2\,ke.
The charge collection distribution obtained with not irradiated modules, biased at a voltage of 150\,V and kept at a temperature of $+20\,^{\circ}\mathrm{C}$ is reported in \cite{NIM_CiS_ALR}, and the measured most probable value (MPV) is (19$\pm$2)\,ke. An overview of the collected charge measured as a function of the bias voltages  measured by irradiated modules is shown in Figure \ref{fig:MPVall}.\\
 \\
Figure 3 shows the charge collection distribution and the hit-map obtained by a sample irradiated at a fluence of $10^{16}$\,1-MeV\,n$_{\mathrm{eq}}$cm$^{-2}$, biased at 600\,V, and with the front-end chip tuned to a threshold of 2\,ke. The distribution refers to all cluster sizes, and the measured  MPV is (4.4\,+\,0.7\,-\,0.4)\,ke, which is more than twice as high as the front-end threshold.
\begin{figure}[h!]
\centering
\subfigure[]{
\includegraphics[scale=0.4]{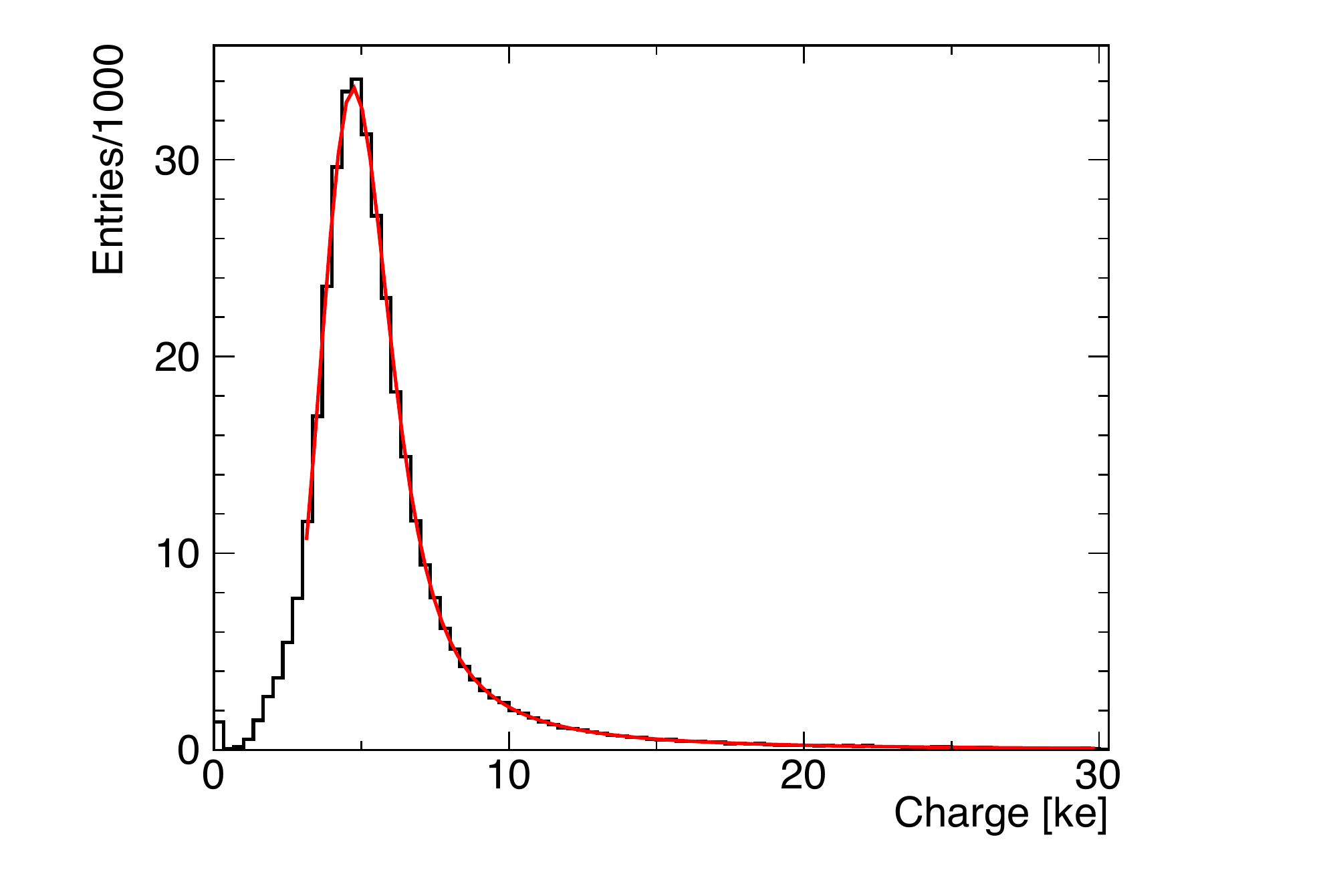}
\label{fig:MPV_th2ke}
}
\subfigure[]{
\includegraphics[scale=0.35]{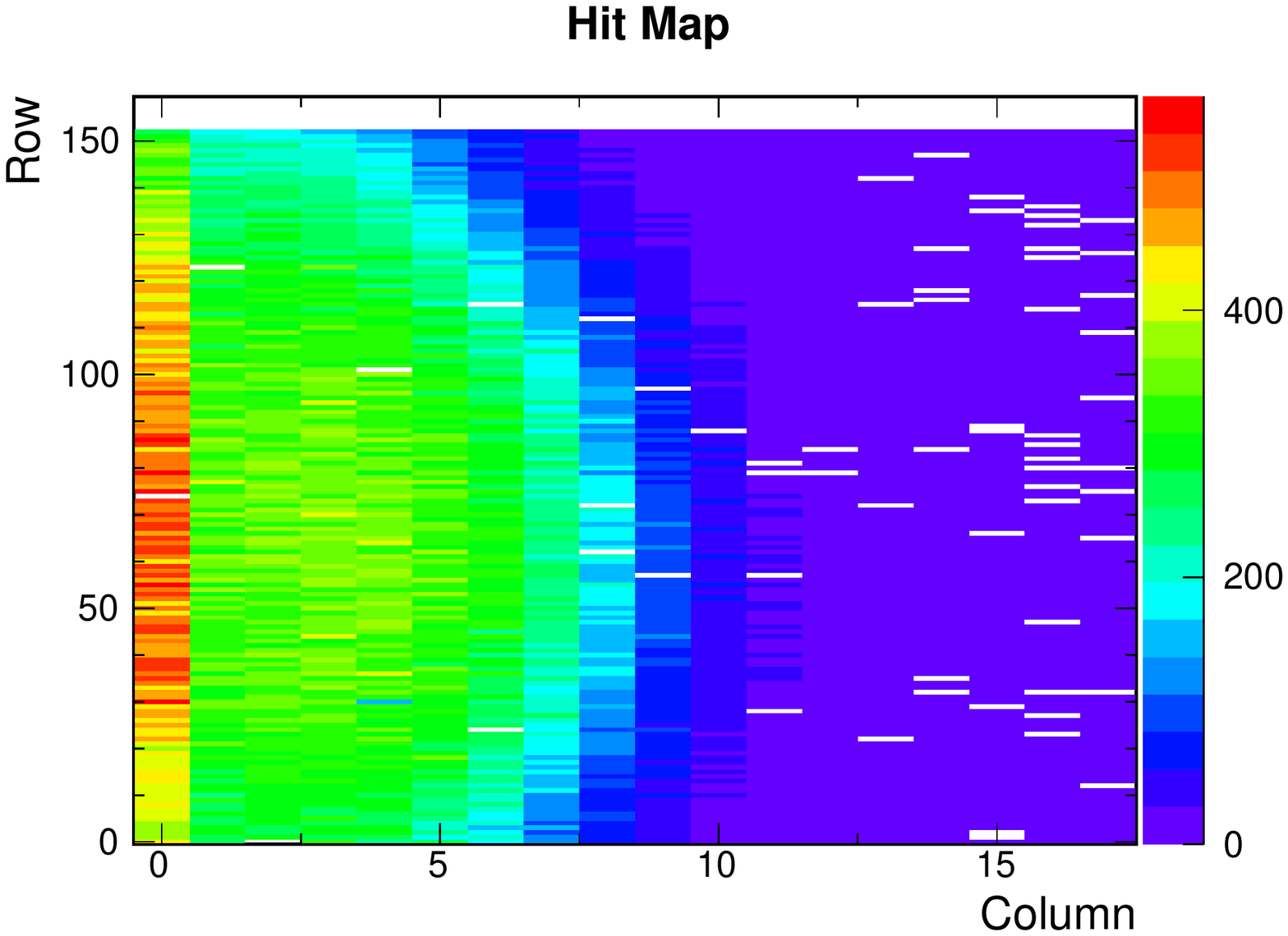}
\label{fig:HitMap}
}
\caption{Charge collection distribution \subref{fig:MPV_th2ke} and hit-map \subref{fig:HitMap} obtained by a module irradiated up to 1\,x\,$10^{16}$\,1-MeV\,n$_{\mathrm{eq}}$cm$^{-2}$.}
\label{fig:MPV_th2ke}
\end{figure}
\\
\indent {\bf Tracking efficiency.} The tracking efficiency has been studied in beam tests. For a not irradiated module it was found to be (99.3$\pm$0.2)\% when setting a front-end threshold of 3.2\,ke, and a bias voltage of 150\,V. For a module irradiated up to 5\,x\,$10^{15}$\,1-MeV\,n$_{\mathrm{eq}}$cm$^{-2}$, the tracking efficiency is (98.6$\pm$0.3)\% with a front-end threshold of 3.2\,ke, and a bias voltage of  600\,V.

\section{Summary and future plans}
\label{sec:summary}

\indent Results of the characterization of n-in-p modules, irradiated up to $10^{16}$\,1-MeV\,n$_{\mathrm{eq}}$cm$^{-2}$, have been presented. The good performance proves the feasibility of employing this technology up to the highest fluence investigated.\\
\indent A new production of n-in-p FE-I4 compatible sensors has been recently completed at CiS on 4". Furthermore, following the upgrade of the CiS production line to process 6" wafers, an additional production on 6" Fz p-type material is foreseen for 1-chip and 4-chip FE-I4  \cite{FEI4} sensors.\\
\indent The work presented was partially performed in the framework of the CERN RD50 Collaboration and the European Commission under FP7 Research Infrastructures project AIDA, grant agreement no. 262025.

\end{document}